\documentclass{jfm}
\usepackage{physics}
\usepackage{graphicx}
\usepackage{newtxtext}
\usepackage{newtxmath}
\PassOptionsToPackage{x11names}{xcolor}
\usepackage{newtxtext}
\usepackage{newtxmath}

\usepackage{caption}
\usepackage{subcaption}
\usepackage{natbib}
\usepackage{hyperref}

\hypersetup{
    colorlinks = true,
    urlcolor   = blue,
    citecolor  = black,
}
\newcommand{\DD}{\mathrm{D}}

\title{Memory effects in wave-induced microplastic transport}
\author{Mary Eby\aff{1} 
\and Cathal Cummins\aff{1}$^,$\aff{2} \corresp{\email{c.cummins@hw.ac.uk}}}
\affiliation{\aff{1}Maxwell Institute for Mathematical Sciences, Department of Mathematics, Heriot-Watt University, Edinburgh EH14 4AS, UK
\aff{2}Institute for Sustainable Built Environment, School of Energy, Geoscience, Infrastructure and Society, Heriot-Watt University, Edinburgh EH14 4AS, UK}

\begin{document}

\maketitle

%\runninglinenumbers

\begin{abstract}
Microplastics are transported by ocean surface waves in ways that depart significantly from the Stokes drift of fluid parcels, and accurate modeling of this transport requires accounting for forces beyond linear drag. Existing modeling of microplastic transport often neglect the Basset-Boussinesq history force, effectively limiting their use to the smallest particle sizes. Here, we extend the applicability of these models by implementing the history term with a multistep integration scheme, allowing us to capture the transport of larger microplastics in linear surface waves of arbitrary depth. We quantify when the Basset-Boussinesq history force significantly affects microplastic transport by surface gravity waves. We show that memory effects become the leading-order horizontal drag once $S=St/\gamma^2$ exceeds a critical value $S\approx 0.25$, where $St$ is the Stokes number and $\gamma$ is the density ratio of the particle and the fluid. The corresponding critical $St$ number is found to be a factor of about three smaller than that given by classical inertial estimates that neglect history effects. These results help provide regime maps that can be used to indicate when history effects can be safely neglected. Our simulations also reveal that history effects significantly increase horizontal transport distances and enhance orbit shearing of particle ensembles.

\end{abstract}
% {\bf MSC Codes }  {\it(Optional)} Please enter your MSC Codes here

\section{Introduction} \label{sec:intro}
The transport of small inertial particles by a fluid flow is an important process in many industrial and environmental phenomena \citep{Daitche2014}, including the transport of marine microplastic debris. Currently, there are over 5 trillion pieces of plastic debris in the global oceans, and approximately 20 million tons of plastics are added each year \citep{Salvador2017, Borrelle2020}. While larger plastic debris (diameter $> 25$mm) constitute the majority of initially buoyant plastics in the ocean, these plastics can degrade into smaller fragments over time due to processes such as biodegradation, photodegradation, and stresses from turbulent flows \citep{Browne2007, Sutherland2023}. As a result, microplastics, or plastic particles with diameter $< 5$mm, are a significant component of marine plastic pollution. Many researchers have approached the study of microplastic transport using oceanographic \citep{Sebille2020} and laboratory \citep{Sutherland2021} observations. However, it is estimated that the vast majority of marine microplastics do not remain at the surface, making it difficult to determine the ultimate fate of the debris \citep{Sebille2015, Sutherland2023}. 

Mathematical models based on the Maxey-Riley equation provide a predictive approach to studying microplastic transport. These models are applicable across a range of parametric regimes, depending on factors such as particle size, fluid velocity, and wave steepness. Modeling efforts have focused on regimes characterized by negligible history effects, linear or nonlinear wave fields, and the presence or absence of wave breaking. Such models can provide valuable insight into inertial particle dynamics, but they are typically restricted to narrow parameter ranges and sometimes omit key physical effects such as memory forces. For example, \citet{Haller2008} used a model based on the Maxey-Riley equation to explore the dynamics of small, inertial particles in unsteady fluid flows, omitting the Basset-Boussinesq history force (hereafter referred to as ``the history force,'' ``history effects,'' or ``memory effects'') on the assumption of sufficiently small particle diameter. Extending singular perturbation techniques to unsteady flows, the authors derived a general asymptotic form of inertial particle motion. \citet{Santamaria2013} examined the horizontal and vertical Stokes drift of small particles transported by linear surface waves, again omitting history effects under the assumption of small Stokes numbers. Their model assumes small wave steepness and infinite water depth. Although the analytical solutions provided are valid under these assumptions, the model's applicability may be limited if history effects become important.

More recently, \citet{Dibenedetto2022} explored both linear and non-linear drag regimes for the transport of small, inertial, negatively buoyant particles in linear waves, again omitting the history force. The inclusion of the non-linear Schiller-Neumann drag model allows for both low and transitional particle Reynolds numbers, thus expanding the scope of previous models for microplastic transport. However, like earlier works, this model assumes sufficiently small particles to justify omitting the history force, which restricts its relevance to only part of the microplastic size spectrum (see Table~\ref{tab:dimensional_parameters}).

Some studies have included the Basset-Boussinesq history force explicitly. \citet{Daitche2013} developed high-order quadrature schemes to numerically solve the Maxey-Riley equation with history effects and demonstrated its accuracy with comparison to analytical solutions. However, buoyancy was not considered in that formulation. Separately, \citet{Prasath2019} formulated the Maxey-Riley equation with memory effects as a time-dependent Robin boundary condition of a one-dimensional diffusion equation and solved it using the unified transform method. Several example applications are also included, such as a sedimenting particle, a particle in an oscillatory flow, and a particle in a Couette flow. In the example of a particle settling under gravity, \citet{Prasath2019} shows that the particle velocity decays exponentially when history effects are neglected, but decays as $ t^{-3/2} $ when the history force is included. 

It is unclear then, under what conditions can the history force be neglected in models of microplastic transport, and when does it fundamentally alter the particle behavior? Indeed, we should clarify: what range of $St$ corresponds to "negligible history effects" or "non-negligible history effects" in the context of wave-induced microplastic transport? Answering these questions is crucial because neglecting history effects has been the default assumption in most wave-particle models, yet microplastics span a wide range of sizes and densities for which this approximation may not be entirely valid. We aim to close this gap by quantifying when history effects become dynamically important and to assess their implications for transport in wave-driven flows. We address this by computing $\chi$, the ratio of the magnitude of the history force to that of the Stokes drag, over a parametric range relevant to microplastic transport. This will provide operational thresholds for researchers to use when modeling the transport of large microplastics in ocean waves, allowing us to state the value of $St$ at which history effects are ``negligible,'' ``non-negligible,'' or ``dominant.''

In \textsection \ref{sec:model}, we present the equations defining our mathematical model of a particle transported by a horizontally propagating surface gravity wave, providing the equation of motion and analytical expressions of the flow field before producing the dimensionless forms of the equations. Then, we discuss the details of the model implementation, and the inclusion of the Basset-Boussinesq history force. We then present, in \textsection \ref{sec:results}, numerical results for the simulation of microplastic particles transported by linear ocean waves, exploring the impact of history effects on the movement of particles of various sizes and densities. Conclusions are presented in \textsection \ref{sec:conclusion}.

\section{Mathematical modeling} \label{sec:model}
\subsection{Inertial particle motion in surface gravity waves} \label{subsec:equations}
\begin{figure}
    \centering
    \includegraphics[width=0.8\textwidth]{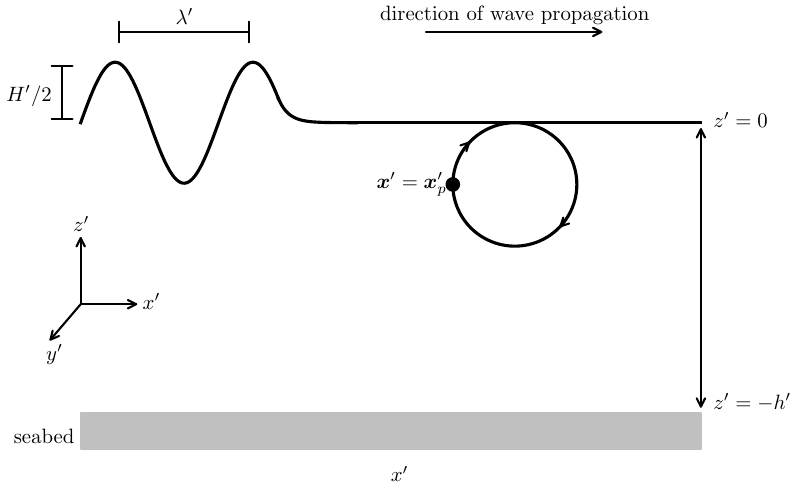}
    \caption{Transport of a microplastic particle by a surface gravity wave, where the particle is located at a wave period endpoint as defined in \textsection{\ref{subsec:param_variation}}}
    \label{fig:seabed}
\end{figure}

We consider a rigid, spherical, inertial particle moving through a surface gravity wave of an incompressible fluid as described by linear wave theory. We assume the system, illustrated in Figure \ref{fig:seabed}, is subject to beam seas along the $y'$-axis, and so we take the cross-section to obtain a 2-D system on the $x'$-$z'$ plane. For the fluid velocity $ \vb* u' = \langle u', w' \rangle $, we have,
\begin{eqnarray}
    u'(\vb*{x}', t') &=& \omega'H' \frac{\cosh(k'(z' + h'))}{2\sinh(k'h')} \cos(k'x' - \omega' t'),\label{eq:dim_u}\\
    w'(\vb*{x}', t') &=& \omega'H' \frac{\sinh(k'(z' + h'))}{2\sinh(k'h')} \sin(k'x' - \omega' t'),\label{eq:dim_w}
\end{eqnarray}
in which $ u' $ is the velocity in the horizontal $ x' $-direction, $ w' $ is the velocity in the vertical $ z' $-direction, and $ h' $ is the water depth. Here and throughout, a prime denotes a dimensional quantity. The angular frequency $\omega'$ of the incident surface gravity wave is related to the wave number $k' = 2\upi / \lambda' $ by the dispersion relation \( \omega' = \sqrt{g' k' \tanh{(k'h')}} \), where $ \lambda' $ is the wavelength, $ H'/2 $ is the wave amplitude, and $g'$ is the acceleration due to gravity.

The equation of motion for the system depicted in Figure \ref{fig:seabed} is the Maxey-Riley equation \citep{Haller2008, Babiano2000, Maxey1983},
\begin{eqnarray} \label{eq:dim_M-R}
    \rho'_p \vb*{\dot{v}}' &=& \rho'_f \frac{\DD \vb*{u}'}{\DD t'}
        + \pqty{\rho'_p - \rho'_f} \vb*{g}'
        - \frac{9 \nu' \rho'_f}{2a^{\prime 2}} \pqty{\vb*{v}' - \vb*{u}'}
        - \frac{\rho'_f}{2} \bqty{\vb*{\dot{v}}' - \frac{\DD \vb*{u}'}{\DD t'}} \nonumber\\&\quad&
        - \frac{9 \rho'_f}{2a'} \sqrt{\frac{\nu'}{\upi}} \int_0^{t'} \frac{\vb*{\dot{v}}'(s') - \vb*{\dot{u}}'(s')}{\sqrt{t' - s'}} \dd s'.
\end{eqnarray}
in which $ \DD \vb*{u}' / \DD t' = \partial_{t'} \vb*{u}' + \vb*{u}' \cdot \nabla' \vb*{u}' $ is the material derivative evaluated at the location of the particle, $ \vb*{v}'(t') = \vb*{\dot{x}}'_p(t') $ is the particle velocity, $ \vb*{x}'_p(t') = \langle x'_p(t'), z'_p(t') \rangle $ is the particle position, $ \vb* u'(\vb*{x}'_p, t') = \langle u'(\vb*{x}'_p, t'), w'(\vb*{x}'_p, t') \rangle $ is the fluid velocity at the location of the particle, $ t' $ is the time, $ \rho'_p $ and $ \rho'_f $ are the particle and fluid densities respectively, $ a' $ is the particle radius, $ \nu' $ is the kinematic viscosity, and gravity $ \vb*{g}' $ is a constant vector. For the flow under consideration, the Fax\'en corrections are equivalently zero.

To make \eqref{eq:dim_u}, \eqref{eq:dim_w} and \eqref{eq:dim_M-R} dimensionless, we use the characteristic scales $ L' $, $ U' $, and $ T' = L' / U' $ to non-dimensionalize space, velocity, and time, respectively. Defining the length scale as $ L' = 1 / k' $, we have $ \vb* x = k' \vb*{x}' $, $ \vb* x_p = k' \vb*{x}_p' $, $ h = k'h' $, and $ \epsilon = k'H'/2 $, where $ \epsilon $ is the wave steepness. We set the velocity scale to be equal to the wave celerity $U'=\omega'/k'$,  providing $ \vb* u = k' \vb*{u}' / \omega' $ and $ \vb* v = k' \vb*{v}' / \omega' $. To maintain the relationship $ T' = L' / U' $, we define $T' = 1 / \omega'$, thus we have $t = \omega' t'$. We also note that the relaxation timescale of the particle is given by $\tau_p'=2 a'^2\rho_p' / (9\rho_f'  \nu')$. Next, we introduce three dimensionless parameters: the density ratio, $ R = 2 \rho'_f/(\rho'_f + 2 \rho'_p) $, the Stokes number, $St =\omega' \tau_p'$,  and the Froude number $Fr = \omega' / \sqrt{g'k'} $, which is approximately equal to 1 for all cases explored throughout this work unless otherwise stated.

We present the governing equations using the density-independent Stokes number $\widehat{St}$, which is related to the standard Stokes number via $St=\gamma \widehat{St} $, where $\gamma =\rho'_p/\rho'_f=1/R-1/2$. We adopt $\widehat{St}$ to enable direct comparison with existing studies (including the original formulation by Maxey and Riley). In reporting thresholds (based on the relative amplitudes of the history to Stokes drag forces), we will use the parameter $S=\widehat{St}/\gamma$, as it provides an approximately uniform threshold across wave conditions. Lastly, using these scalings we obtain  $ \vb* g = k'\vb*{g}' / \omega'^2 $, equivalently $ \vb* g = 1 / Fr^2 $. Then, the dimensionless form of the Maxey-Riley equation is, 

\begin{equation} \label{eq:M-R}
    \vb*{\dot{v}} =  \overbrace{\frac{3R}{2} \frac{\DD \vb* u}{\DD t}}^{\text{inertial forces}} +  \overbrace{\pqty{1 - \frac{3R}{2}} \vb* g}^{\text{gravity}} -  \overbrace{\frac{R}{\widehat{St}} \pqty{\vb* v - \vb* u}}^{\text{Stokes drag}} -  \overbrace{\sqrt{\frac{9}{2 \upi}} \frac{R}{\sqrt{\widehat{St}}} \int_0^t \frac{1}{\sqrt{t - s}} \dv{s} \bqty{\vb* v - \vb* u} \dd s}^{\text{Basset-Boussinesq history force}},
\end{equation}
where terms on the right-hand side correspond to: the inertial forces, the buoyancy/gravitational force, the Stokes drag, and the Basset-Boussinesq history force. The dimensionless expressions for the components of the fluid velocity are,
\begin{eqnarray} 
    u(\vb* x, t) &=& \epsilon\frac{\cosh(z + h)}{\sinh(h)} \cos(x - t),\label{eq:u}\\
    w(\vb* x, t) &=& \epsilon\frac{\sinh(z + h)}{\sinh(h)} \sin(x - t),\label{eq:w}
\end{eqnarray}
where the system is characterized by the Stokes number, density ratio, water depth, and wave steepness, which we assume to be small ($ \epsilon \ll \tanh{(h)} $).
\subsection{Treatment of the Basset-Boussinesq history force} \label{subsec:history}
In many models of inertial particle transport, the Basset-Boussinesq history force is omitted on the grounds that the Stokes number is small ($St \ll 1$), thereby simplifying the system. Its inclusion transforms the problem into an integro-differential equation and introduces numerical difficulties arising from the singular kernel in the history integral \citep{Daitche2015}. Attempts to directly compute numerical solutions, rather than approximating them, incurs a high computational cost. For each time step, the history force must be re-computed, as the integral within the history term is evaluated over all previous time steps. This process of re-computing the history term for every time step becomes increasingly computationally expensive as $ t $ grows.

To address these issues in our implementation, we compute the singular kernel analytically by employing a quadrature scheme first constructed by \citet{Daitche2013} and the corresponding multi-step integration scheme, modified to include the buoyancy force. To apply the quadrature scheme, we rewrite the Maxey-Riley equation \eqref{eq:M-R} to include the expression for the general kernel $ K(t - s) = 1/\sqrt{t-s} $, obtaining,
\begin{equation} \label{eq:quad_scheme_M-R}
    \vb*{\dot{v}} = 
    \frac{3R}{2} \frac{\DD \vb* u}{\DD t}
    + \pqty{1 - \frac{3R}{2}} \vb* g 
    - \frac{R}{\widehat{St}} \pqty{\vb* v - \vb* u} 
    - \sqrt{\frac{9}{2 \upi}} \frac{R}{\sqrt{\widehat{St}}} 
    \dv{t} \int_0^t \dd s \: K(t - s) \pqty{\vb* v - \vb* u}.
\end{equation}
which corresponds to equation (5) in \citep{Daitche2013}. From \eqref{eq:quad_scheme_M-R}, the construction as outlined in \citep{Daitche2013} may be followed to achieve arbitrarily high order integration schemes, though a third order scheme is used throughout this work.

\subsection{Comments on the implementation} \label{subsec:comments}
To decrease the size of the errors incurred in the initial time steps of the integration scheme as described in \citet{Daitche2013}, we begin the integration using smaller time steps $ \sim\Delta t^2 / \sqrt{2}$ with lower order schemes, switching to the full-sized time steps with the third order scheme when applicable. To test the accuracy of the model, especially in the early time steps of the integration scheme, we reproduced results from \citet{Daitche2013} for a rigid rotating particle with an analytical solution. Reproducing the corresponding error analysis provided a relative error on the order of $ 10^{-6} $ using the third order integration scheme. Regarding the timestep size in relation to the relative error, we note that depending on the chosen timestep size, the value of the history force at time $ t = 0 $ may not be zero, but converges to zero as the timestep size goes to zero.

There are two features of the implementation with regards to computational memory costs that are worth mentioning. First, to minimize the memory required to store the pre-computed values of the coefficients of the weighted sum in the quadrature scheme, the values were computed with quad precision (128-bit floating point format) and stored with double precision (64-bit floating point format). However, for long timescales or very small time steps, the implementation causes memory errors on standard computing systems, such as personal computers. A potential improvement on this method is to use a Taylor expansion of the coefficients in the quadrature scheme, as demonstrated in \citet{Gong2023}, to relax the requirement of quad precision.

\subsection{Computation of $\chi$} 
In order to compare the relative amplitude of the drag forces, we evaluate the hydrodynamic forces appearing on the right-hand side of the Maxey-Riley equation \eqref{eq:M-R}, using the numerically computed particle trajectory as input. This yields time series for the Stokes drag and history forces:
\[
\mathbf{F}_{\mathrm{Stokes}}(t)=-  \frac{R}{\widehat{St}} \pqty{\vb* v - \vb* u};
\quad
\mathbf{F}_{\mathrm{hist}}(t)=-  \sqrt{\frac{9}{2 \upi}} \frac{R}{\sqrt{\widehat{St}}} \int_0^t \frac{1}{\sqrt{t - s}} \dv{s} \bqty{\vb* v - \vb* u} \dd s
\]
To characterize the behavior of these forces, we fit each force signal with a function of the form:
\begin{equation} \label{eq:curve_fit}
    A e^{-\delta t} \sin{(\omega t + \phi)} + \text{offset},
\end{equation}
to the $x$- and $z$-components of $\mathbf{F}_{\mathrm{Stokes}}(t)$ and $\mathbf{F}_{\mathrm{hist}}(t)$, where $A$ is the amplitude, $\delta$ is the decay rate, $\omega$ is the angular frequency equivalent to the angular frequency of the velocity field, and $\phi$ is the phase angle relative to the phase angle of the particle velocity. The offset is also included, as it becomes important for certain wave conditions. After the curve fitting, we isolate the variables $A$, $\delta$, $\omega$, $\phi$, and the offset in \eqref{eq:curve_fit}, which correspond to different characteristics of the signal. In order to determine the relative contribution of the history forces, we focus on the forces in the $x$-direction only and form the ratio of the amplitudes of the forces in the $x$-direction:
\begin{equation}
\chi = \frac{A_{\mathrm{hist},x}}{A_{\mathrm{Stokes},x}}.
\label{eq:chi-def}
\end{equation}
Before we provide our results, we first estimate the relative magnitude of these different forces using asymptotic approximations, considering the Maxey-Riley equation \eqref{eq:M-R} in the limit of small $ \widehat{St} /R$. In this limit, the relative velocity may be approximated using the leading-order inertial correction \citep{Haller2008}:
\begin{equation}
\vb*{v} - \vb*{u} \approx \frac{\widehat{St}}{R} \left( \frac{3R}{2} - 1 \right) \left( \frac{D \vb*{u}}{Dt} - \vb*{g} \right)+O\left(\left( \frac{\widehat{St}}{R}\right)^2\right).\label{eq:inertial-corr}
\end{equation}
This approximation is strictly valid only when history effects are neglected, but it provides an \emph{a priori} estimate of when history effects can be neglected. Substituting the inertial correction into the drag terms yields approximate horizontal scalings:
\[A_{\mathrm{Stokes},x}^{\mathrm{approx.}}\sim \beta  \epsilon; \quad A_{\mathrm{hist},x}^{\mathrm{approx.}}\sim \beta \epsilon \sqrt{9 \widehat{St} / 2\upi},\]
where $ \beta := {3R}/{2} - 1  $.  This inertial approximation gives 
\begin{equation}
\label{eq:chi}
    \chi^{\mathrm{approx.}}\sim \sqrt{9\widehat{St} / 2\pi},
\end{equation} 
which indicates that history force overtakes Stokes drag ($\chi>1$) when $ \widehat{St} > {2\pi}/{9}  \approx 0.70$. 

We note that the formal small-particle limit used in the original derivation of the Maxey-Riley equation requires $\widehat{St}\ll 2/9$.
\section{Results} \label{sec:results}
Here, we investigate the contribution of history effects to the transport of particles in surface gravity waves. Since this work is focused on ocean wave-induced microplastic transport, we select simulation conditions based on the parameter ranges relevant to microplastics. The parametric ranges applicable to microplastic particles in ocean waves are outlined in Table \ref{tab:dimensional_parameters}. The corresponding dimensionless parameters are then varied to explore the full parameter space. 

Simulating particle transport using conditions chosen in accordance with the ranges outlined in Table \ref{tab:dimensional_parameters}, we investigate how various forces influence the motion of the particle, focusing especially on the history force. Further, we examine history effects on particles of different sizes and densities, and vary the depth of the water through which the particles are transported. These variations in the particle size, particle density, and water depth, along with the assumption of small wave steepness, correspond to the dimensionless parameters which characterize the system as described in \textsection \ref{sec:intro}. We note that for simulations with a particle Reynolds number in the transitional range, the Stokes drag term in \eqref{eq:M-R} should be modified to allow for nonlinear drag, such as the Schiller-Neumann drag model described in \citet{Dibenedetto2022}. The scope of this paper, however, is focused on the effect of the Basset-Boussinesq history force.
\begin{table}
  \begin{center}
\def~{\hphantom{0}}
  \begin{tabular}{lccc}
    wavelength & $ 1.5\text{m} \leq \lambda' \leq 10^3 $m\\
    wavenumber & $ \upi / 500\mathrm{m}^{-1} \leq k' \leq 4\upi / 3 \mathrm{m}^{-1} $\\
    depth & $ 0 < h' \leq 10,984\text{m}\; (3,682 $m average)\\
    particle radius & $ 0.5\mu\text{m} \leq a' \leq 2.5\text{mm} $\\
    kinematic viscosity & $ \nu' = 10^{-6} \text{m}^2/\text{s} $\\
    particle density & $ 0.85\mathrm{g}/\mathrm{cm}^3 \leq \rho_p' \leq 1.41\mathrm{g}/\mathrm{cm}^3 $\\
    fluid density & $ 0.943\mathrm{g}/\mathrm{cm}^3 \leq \rho_f' \leq 1.0962\mathrm{g}/\mathrm{cm}^3 $\\
    Stokes number & $ 0 < \widehat{St} \leq 8.9032 $\\
    wave steepness & $ 0 < \epsilon \leq 0.1 $\\
    density ratio & $ 0.50120 \leq R \leq 0.81314 $\\
  \end{tabular}
  \caption{Parametric ranges relevant to microplastic particles transported by ocean waves \citep{Calvert2021, Toffoli2017, Jambeck2015, Dibenedetto2022, Sutherland2023, Sebille2020, Stocchino2019, Sharqawy2010, Nayar2016, Gardner2014}}
  \label{tab:dimensional_parameters}
  \end{center}
\end{table}
\subsection{Analysis of forces acting on the particle} \label{subsec:forces}
The full Maxey-Riley equation  includes four force contributions: inertial, gravitational, Stokes drag, and memory effects via the Basset-Boussinesq drag term, as indicated in \eqref{eq:M-R}. Unlike Stokes drag, which depends instantaneously on the particle's velocity relative to the surrounding fluid, the history force depends on the entire past relative acceleration of the particle. As a result, it behaves qualitatively differently from the other hydrodynamic forces, and this distinction has significant implications for particle transport.

\begin{figure}
\centering
\includegraphics{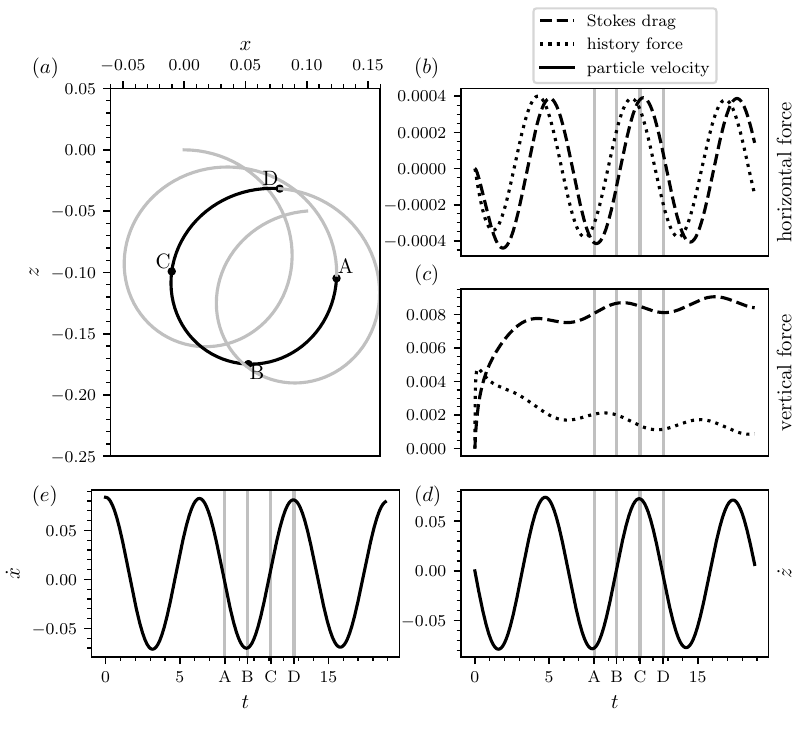}
\caption{The Stokes drag and history force contributing to the movement of a small, negatively buoyant particle in a deep water wave, with $ S = 0.2 $ $ (St = 0.20611) $, $ R = 0.66 $ and $ \epsilon = 2\upi / 75 $. The particle trajectory $(a)$ is shown in the $x$-$z$ plane, with four points in time A-D at different quarters of the central particle orbit, highlighted by the vertical lines in the remaining plots. The forces are split into their horizontal $(b)$ and vertical $(c)$ components. The particle velocity is similarly split into its vertical $(d)$ and horizontal $(e)$ components.}
\label{fig:forces}
\end{figure}
We now turn to a representative simulation of a negatively buoyant microplastic particle in a linear deep-water surface gravity wave ($S = 0.2$, $R = 0.66$, and $\epsilon = 2\upi / 75$). These parameter values lie within the ranges in Table \ref{tab:dimensional_parameters} relevant to microplastics. Figure \ref{fig:forces} shows the individual force components over one oscillation cycle, decomposed into horizontal and vertical contributions.
The particle's horizontal motion is oscillatory and in phase with the underlying wave field while its vertical motion is governed by slow gravitational settling in agreement with the inertial correction \eqref{eq:inertial-corr}. Points A, B, C, and D mark four key stages in the horizontal cycle, each spaced approximately a quarter-period apart. At points A and C, the particle reaches its maximum horizontal displacement and reverses direction, so its horizontal velocity vanishes. At B and D, the particle reaches its maximum horizontal speed; first against, then with the wave direction.

At point A, the particle reaches its furthest point in the direction of wave propagation and begins to reverse. Its horizontal velocity is zero there, and its acceleration is large and negative. At this point, the Stokes drag is at a local minimum, and the history force is increasing in response to the earlier acceleration. Vertically, gravity remains dominant (it scales as $\sim \beta/Fr^2$), but the Stokes drag has overtaken the history force as the largest dynamic contribution since the constant gravity contribution to the history integral disappears from the history force so it remains $ \sim \beta \epsilon \sqrt{9 \widehat{St} / 2\upi} $ in the vertical direction while Stokes drag eventually scales as $\beta/Fr^2$ to match gravity as expected from  \eqref{eq:inertial-corr}.

At point B, the particle has reached maximum speed moving in the opposite direction to the wave propagation. The relative velocity in the horizontal direction is momentarily zero (since the fluid velocity matches that of the particle), so the Stokes drag crosses zero. The horizontal history force nears a local maximum, having accumulated the earlier relative accelerations. Vertically, the Stokes drag continues to increase with settling speed, while the history force decays.

Point C mirrors point A: the particle reverses again, this time beginning to move in the wave direction. The horizontal velocity is again zero, the particle acceleration is strongly positive, and the horizontal Stokes drag reaches a local maximum. In the vertical direction, the force balance is similar to that at points A and B: gravity dominates, with Stokes drag the largest dynamic term. The history force in the vertical direction continues to decrease.

At point D, the particle is moving fastest in the direction of wave propagation. Like at B, the relative velocity in the horizontal direction is zero, so the Stokes drag vanishes. The history force in the horizontal direction now reaches a negative peak. Again, the vertical behavior is qualitatively the same as at previous points: the Stokes drag continues to grow as settling speed increases, while the history force continues to decay.

This shows the key differences in the phase and amplitude of the Stokes and Basset drag forces. In the horizontal direction, where no body force acts, the balance between inertial, history, and Stokes drag forces drives the oscillatory motion. In this case, the history force and Stokes drag contribute approximately equally ($\chi \approx 0.9$) to the horizontal drag term across the cycle. This behavior is consistent with the qualitative trends predicted by the inertial approximation of the particle’s relative velocity. In the vertical direction, gravity remains dominant, but the relative contributions of the dynamic forces evolve: the history force initially exceeds the Stokes drag, before becoming subdominant as the particle continues to settle. Hence, memory effects can be significant contributors to the horizontal drag force present in microplastic transport; they alter the magnitude of the forces involved and they affect the timing of particle responses. They introduce phase lags that can influence long-term dispersion and settling behaviors, which are not captured by models relying on Stokes/instantaneous drag alone. 

Since microplastic particles in linear ocean waves are both oscillating and sedimenting, we draw comparisons from Figure \ref{fig:forces} to existing examples of particles in sedimenting and oscillatory flows, as in \citet{Prasath2019}. The sedimenting particles illustrated in Figure 5 of \citet{Prasath2019} demonstrate that the inclusion of history effects causes the particles to reach their terminal velocity much more slowly than if history effects were omitted. Conversely, the oscillating particles in Figure 7 of \citet{Prasath2019} indicate that the particles reach their final periodic state more quickly with history effects, and further, there exists a phase difference between oscillating particles with and without history effects on long timescales.

Since there are no gravitational effects acting in the horizontal direction, the oscillatory effects are most evident in the evolution of the horizontal forces over time, shown in Figure \ref{fig:forces}. We observed that the magnitude of the history force in this direction is comparable to the other forces, is a significant contributor to the drag force, and is not in phase with the Stokes drag. The combination of these characteristics implies a phase difference between the horizontal trajectories of particles with history effects and those without, as observed in \citet{Prasath2019}. In contrast, the vertical forces shown in Figure~\ref{fig:forces} are dominated by gravitational effects (since they scale as $\beta/Fr^2$), and thus the characteristics of sedimentation are visible. Due to the magnitude of the constant gravitational force, the history force does not contribute as significantly as it does in the horizontal direction. However, it should be noted that the magnitude of the history force is still greater than that of the Stokes drag initially, and decays over time.

\begin{figure}
    \centering
\includegraphics{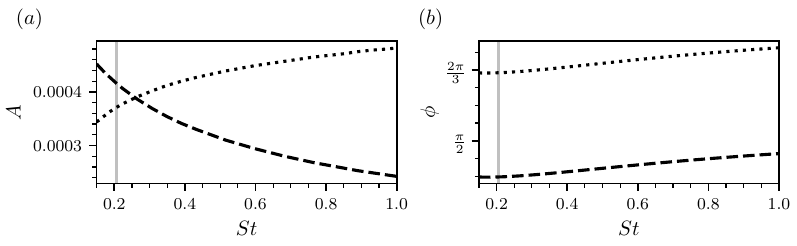}
    \caption{The Stokes number vs. the signal amplitude $A$ $(a)$, and phase angle $\phi$ $(b)$ of the horizontal components of the Stokes drag (dashed) and history force (dotted) acting on a particle with $ R = 0.66 $ and $\epsilon = 2\upi / 75$. The vertical line indicates the position of the Stokes number used in the simulation displayed in Figure \ref{fig:forces}}
    \label{fig:horizontal_forces}
\end{figure}

Now that we have seen an example in which history effects significantly influence microplastic transport, we proceed to explore how the importance of the history force varies with system parameters. We begin by varying the Stokes number, which controls particle inertia and is directly related to particle size. In Figure \ref{fig:horizontal_forces}, we focus specifically on the amplitude and phase angle of the forces appearing in \eqref{eq:curve_fit}.
The intersection of the amplitude of the history force and the Stokes drag shown in Figure \ref{fig:horizontal_forces}$(a)$ demonstrates that the Stokes drag is only more significant to the horizontal motion of the particle than the history force for very small Stokes numbers, and thus very small particles. For the specified density ratio $R=0.66$, only for particles with ${St}<0.25$ is the Stokes drag greater than the history force. The vertical line in Figure \ref{fig:horizontal_forces} corresponds to the example illustrated in Figure \ref{fig:forces}.

In Figure \ref{fig:horizontal_forces}$(b)$, the difference between the phase angles of the forces are in agreement with the results shown in Figure \ref{fig:forces}, and we remark that the history force is out of phase with the Stokes drag. The observation that the drag forces are out of phase with each other and out of phase with the horizontal velocity supports the findings in \citet{Prasath2019} Figure 7$(b)$, which illustrates that the horizontal velocity of a particle transported with history effects is slightly out of phase with the horizontal velocity of a particle transported without history effects.

To identify the point at which the history force overtakes the Stokes drag, we define $St^*$ as the Stokes number at which the amplitudes of their horizontal components are equal, i.e. $\chi=1$ -- see \eqref{eq:chi-def}. In Figure \ref{fig:horizontal_forces}, for example, $ St^* = 0.25$. The corresponding density-independent critical Stokes number is $\widehat{St}^* = 0.25$. It is sometimes more useful to represent our results in this $\widehat{St}$ because it displays less variability over $R$. We also compute where the history force is 75\% that of the Stokes drag ($\chi=0.75$), and for the case of $R=0.66$, we can estimate $\chi\to 0$ given that it follows $\chi\sim \sqrt{\widehat{St}}$  from equation \eqref{eq:chi}; we find that this approximation holds well in Figure \ref{fig:parameter_map}$(b)$, and is used to classify the Stokes drag dominant regime based on the data emerging from the non-negligible history and history dominant regimes. Over the parametric range considered, we find that our previous ``best guess'' estimate $(\widehat{St}^*=2\pi/9 \approx 0.70)$ significantly overestimates the threshold at which memory effects become important/dominant. 
\subsection{Regime map}
\begin{figure}
    \centering
    \includegraphics{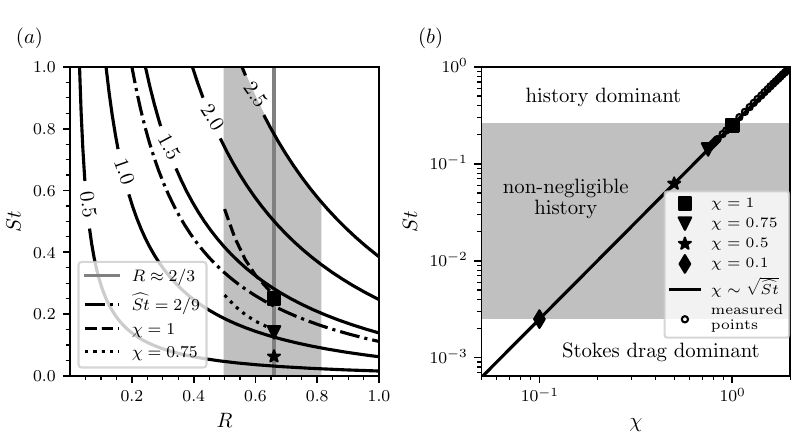}
    \caption{Contours illustrating the particle radius $a'$ computed with varying Stokes number $St$, varying density ratio $R$, and constant wavelength $ \lambda' = 200$m $(a)$, and the Stokes number as a function of the drag force ratio $ \chi $, delineating the different regimes according to the significance of the history force. The vertical line in $(a)$ denotes the approximately neutral buoyancy used in the computation of the data points shown in $(b)$, the dot-dashed line corresponds to the bound for the Maxey-Riley equation, and the dashed and dotted lines display numerically computed values of $St$ and $R$ for the respective values of $\chi$. The shaded area in $(a)$ denotes the region applicable to microplastics. In $(b)$, the line corresponds to the inertial approximation \eqref{eq:inertial-corr}, where $ St \approx \widehat{St} \approx S $ for the approximately neutrally buoyant density ratio. The circular markers indicate numerical results, and the filled markers indicate estimations for notable values of $\chi$.}
    \label{fig:parameter_map}
\end{figure}
In the preceding subsection, we have quantified how history effects modify the dynamics of microplastics for a range of Stokes numbers. However, it is still unclear at what Stokes number can history effects be neglected in place of Stokes drag. Figure \ref{fig:parameter_map} provides a regime map that identifies when history forces are negligible, comparable to, or dominant over Stokes drag. It shows where various ratios of $\chi=0.1, 0.5, 0.75, 1$. We classify these regimes using the parameter $S = \widehat{St}/\gamma$. We find that for fixed values of $\chi$, the quantity $S$ remains approximately constant across the density ratios relevant to microplastics:
\begin{itemize}
    \item \emph{History force is negligible, Stokes-drag regime} ($\chi< 0.1$): the history force contributes less than 10\% of Stokes drag, and corresponds to $S<0.0025$. For a representative deep-water wave (with $\lambda' =200$\,m), then we can estimate that spherical polystyrene particles ($\rho_p'=1050$\,kg m$^{-3}$, $R = 0.656$) of radius $a'<0.14$\,mm in the ocean fall within this regime.
     \item \emph{History force is non-negligible} ($0.1\leq \chi \leq 1$): omitting history forces will introduce appreciable error, and corresponds to $0.0025\leq S\leq 0.25$. For the same wave conditions as above, we can estimate that polystyrene particles of radius $0.14 \,\text{mm} \leq a'\leq 1.4\,\text{mm}$ fall within this regime.
     \item \emph{History force is dominant} ($\chi>1$): the history force exceeds the Stokes drag, and this corresponds to $S>0.25$ over the density ratios of typical microplastics. For the same wave conditions as above, we can estimate that polystyrene particles of radius $1.4\,\text{mm}< a'$ fall within this regime. 
\end{itemize}

In Figure \ref{fig:parameter_map}a, we plot the $\chi = 1$ and $\chi = 0.75$ curves for various density ratios; the $\chi = 0.1$ and $0.5$ thresholds are estimated by fitting a curve $\chi = \kappa \sqrt{\widehat{St}}$ (using the expected behavior from the inertial approximation in \eqref{eq:chi} but with $\kappa\approx 2$) illustrated in  Figure \ref{fig:parameter_map}b, as simulations become computationally expensive at very small $\widehat{St}$. 

Given that previous work to implement a predictive model of inertial particle transport has been conducted either excluding history effects, or without consideration for flows that are both oscillatory and sedimenting, we explore the non-negligible history regime that is valid within the Maxey-Riley framework. Within this regime, we are able to examine the combined effects of oscillation, sedimentation, and large microplastic particle sizes for which history effects are important. While the outlined regimes and the associated parameter map in Figure \ref{fig:parameter_map} will be used to contextualize the results presented in this work, we note that they may also be used retrospectively on existing literature to determine whether the specified Stokes numbers are appropriately small to support the exclusion of history effects. Such an exercise could provide additional context to past results, as the vast majority of examined studies fall outwith the Stokes drag regime.
\subsection{Variation in particle size and density} \label{subsec:param_variation}
We continue our exploration of variations in particle size, focusing now on the horizontal Stokes drift velocity of the particle throughout the water column. The Stokes drift is the net drift in the direction of the wave propagation observed for particles in surface gravity waves \citep{Bremer2017}. As a particle oscillates, it spends more time moving forward under the wave crest than backwards under the wave trough. Under the wave crest, the particle is closer to the water surface where the horizontal velocity has larger magnitude. The fact that these small-scale dynamics result in the globally observable characteristic of horizontal drift make the Stokes drift velocity a useful metric to connect the force-level dynamics observed in Figure \ref{fig:forces} to trends observed over the course of an entire particle trajectory. For simplicity, we begin with a neutrally buoyant particle. An exact solution for the horizontal drift velocity $ u_d $ of a neutrally buoyant particle is defined in \citet{Bremer2017, Ursell1953}, and so we non-dimensionalize the solution using the mappings listed in \textsection\ref{subsec:equations} and employ
\begin{equation} \label{eq:analytical_SD}
    u_d = \frac{\cosh(2(h + z))}{2\sinh^2(h)},
\end{equation}
for comparison to our numerical results for the drift velocity $\bar{\vb* u}$, computed as the displacement of the particle position $ \vb*{x}_p $ over each wave period $T$,
\begin{equation} \label{eq:numerical_drift_vel}
    \bar{\vb* u}_T = \frac{\vb* x_{pT} - \vb* x_{p(T - 1)}}{t_T - t_{T - 1}}.
\end{equation}
The endpoint of a wave period is defined as the point where the horizontal Lagrangian velocity changes from negative to positive, as in \citet{Santamaria2013}. This period endpoint also represents the average vertical position $\bar{z}$ of a particle within its orbit. For an illustration of the locations of these period endpoints in a particle trajectory, see Figure \ref{fig:seabed}. The average vertical position of the particle over the course of its entire trajectory $\bar{\bar{z}}$ is computed by taking the mean of all $\bar{z}$ values.
\begin{figure}
    \centering
    \includegraphics{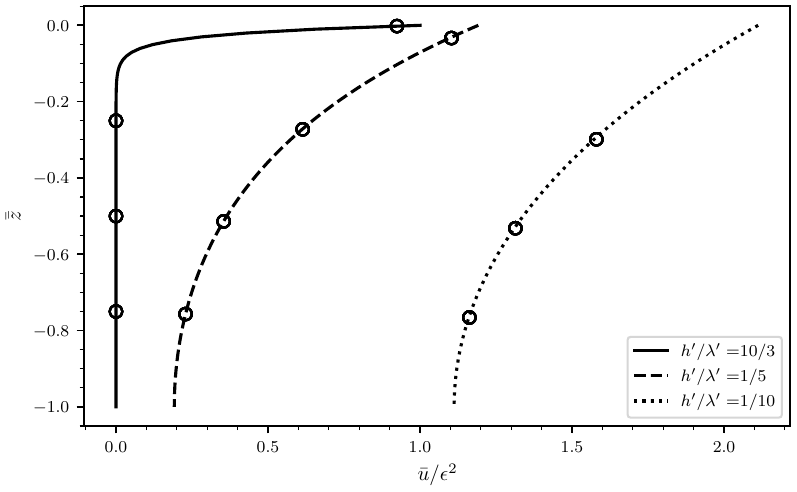}
    \caption{The average vertical position of the particle $\bar{\bar{z}}$ normalized with the water depth $h$ as a function of the period-averaged horizontal component of the normalized Stokes drift velocity $\bar{u}/\epsilon^2$. The particles are neutrally buoyant ($ R = 2 / 3 $) with constant wave steepness $ \epsilon = \upi / 75 $, varying Stokes number, and varying depth. The curves correspond to the analytical solutions for the Stokes drift velocity, and each point corresponds to the numerical solutions averaged over the course of the trajectory of an individual particle. For $ h' / \lambda' = 10/3 $ (the deep water case), $ h = 20\upi / 3 $ and $ Fr \approx 1 $. For $ h' / \lambda' = 1/5 $ (intermediate depth), $ h = 2 \upi / 5 $ and $ Fr \approx 0.92 $. For $ h' / \lambda' = 1/10 $ (the shallow water case), $ h = \upi / 5 $ and $ Fr \approx 0.75 $.  The results showed no dependence on $\widehat St$.}
    \label{fig:varying_depths}
\end{figure}

In Figure \ref{fig:varying_depths}, varying the Stokes number has no effect on the numerical results when $R=2/3$: the curves coincide for all particle sizes, release depths, and whether or not history effects are included. The curves corresponding to the exact solutions from \eqref{eq:analytical_SD}, show that the numerical and analytical solutions agree across all water depths considered in this study. This demonstrates that the history force is negligible for neutrally buoyant particles irrespective of the particle size, which corroborates the drift velocity results from the analytical study in \citet{Santamaria2013} equations (13) and (14).

\begin{figure}
    \centering
    \includegraphics{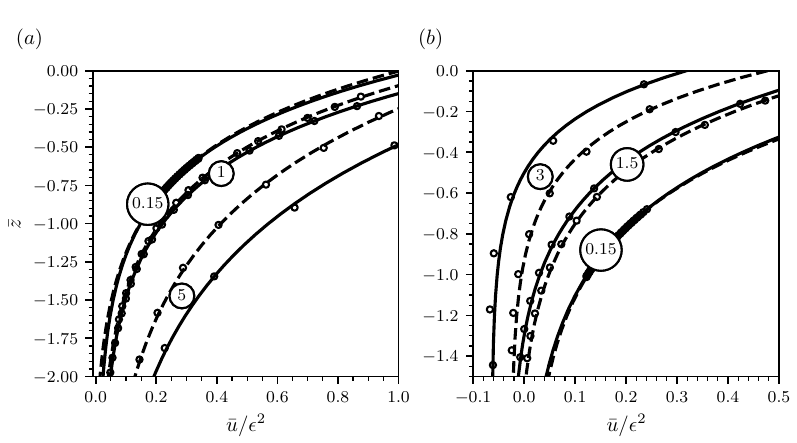}
    \caption{The period-averaged vertical position of a particle $\bar{z}$ moving through deep water waves with $ \epsilon = 2 \upi / 75 $ as a function of the horizontal component of the period-averaged Stokes drift velocity $\bar{u}$ when varying the Stokes number for negatively buoyant particles $(a)$ and positively buoyant particles $(b)$. In both cases, the density ratio was held constant, with $ R = 0.66 $ for the negatively buoyant case and $ R = 0.67\bar{3} $ for the positively buoyant case. Each curve corresponds to a simulation either with (dashed) or without (solid) history effects, with points indicating the numerical solutions averaged over each wave period and circular labels indicating the value of $S$.}
    \label{fig:varying_St}
\end{figure}

Now, we investigate how the drift velocity of positively ($R>2/3$) and negatively ($R<2/3$) buoyant particles is influenced by the Stokes number. We examine the horizontal drift velocity of the particle throughout the water column, with and without history effects. We observe in Figure \ref{fig:varying_St} that as the particle size grows with constant density, the difference between solutions with the history force and those without the history force increases. For the smallest particles simulated in this example, with $ S = 0.15 $, the solutions for the Stokes drift velocity are overlapping throughout the water column, suggesting negligible history effects and tracer-like movement. This indicates that the history force becomes more important for larger particles. Additionally, this feature is exhibited for both positively and negatively buoyant particles, and is substantiated by the results shown in Figure \ref{fig:horizontal_forces}$(a)$, where the amplitude of the history force increases as the Stokes number grows, while the amplitude of the Stokes drag decreases. Both Figure \ref{fig:horizontal_forces}$(a)$ and Figure \ref{fig:varying_St} demonstrate that history effects become more significant as the Stokes number, and thus the particle size, increases.

\begin{figure}
    \centering
    \includegraphics{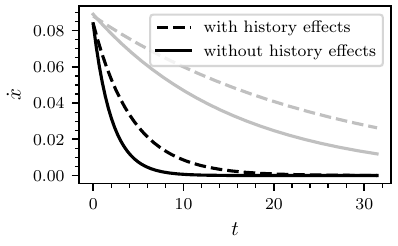}
    \caption{The decay of the horizontal velocity of negatively buoyant particles over time for heavy particles (black) with $ S = 1.5 $ $ (St = 2.7413) $, $ R = 0.54 $ and large particles (grey) with $ S = 10 $ $ (St = 10.305) $, $ R = 0.66 $.}
    \label{fig:velocity_decay}
\end{figure}

\begin{figure}
    \centering
    \includegraphics{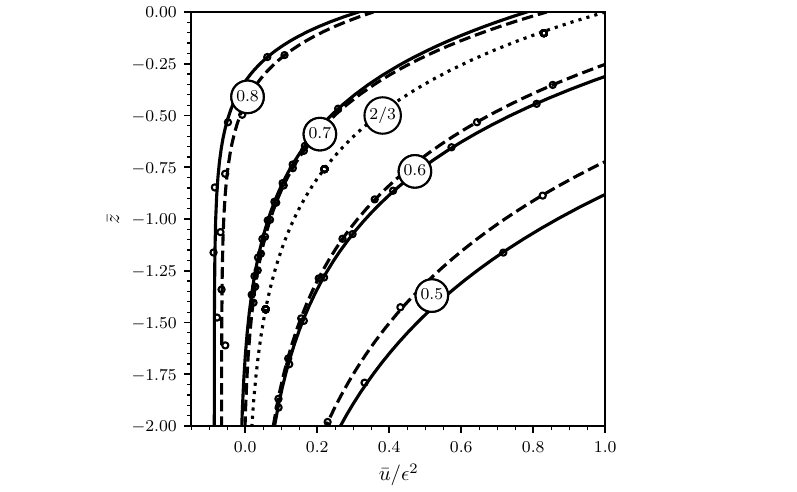}
    \caption{The period-averaged vertical position of a particle $\bar{z}$ moving through deep water waves with $ \epsilon = 2 \upi / 75 $ as a function of the horizontal component of the period-averaged Stokes drift velocity $\bar{u}$ when varying the density ratio with a constant $ \widehat{St} = 0.2 $. Each curve corresponds to a simulation either with (dashed) or without (solid) history effects, with points indicating the numerical solutions averaged over each wave period and circular labels indicating the value of the density ratio. The analytical solution for neutrally buoyant particles (dotted) is also shown, computed with \eqref{eq:analytical_SD}. Numerical solutions for the neutrally buoyant case were computed for particles released from different vertical positions in the water column, as in Figure \ref{fig:varying_depths}}
    \label{fig:varying_R}
\end{figure}

Further, we note that regardless of the particle density, the drift velocity of particles simulated with history effects is closer to that of Stokes drag dominant particles (e.g., $S = 0.15$ curve in Figure \ref{fig:varying_St}$(b)$). In this sense, the history force results in behavior that is more tracer-like, and more faithfully following the underlying wave-induced flow. To quantify this tracer-like behavior, we examine the decay in the amplitude of the horizontal particle velocity over time. By extracting velocity peaks and constructing an envelope (Figure \ref{fig:velocity_decay}), we observe that simulations including the history force display a slower reduction in amplitude compared with those without. Physically, this means that particles remain suspended for longer in the energetic upper water column, where wave-induced velocities are strongest. As a result, it continues to experience significant horizontal accelerations for longer. This finding is consistent with previous analyses of sedimenting or relaxing particles, which also reported slower relaxation when memory effects are included \citep{Prasath2019}.
After examining how varying the Stokes number modifies particle transport (Figure \ref{fig:varying_St}), we now investigate the effect of changing the density ratio. Figure \ref{fig:varying_R} presents the horizontal drift velocity across the water column for a range of $R$, with $\widehat{St}=0.2$ held fixed. As in Figure \ref{fig:varying_depths}, the analytical solution \eqref{eq:analytical_SD} is included for comparison.

We note that the difference between simulations run with and without history effects widens as the density deviates from neutrally buoyant, a feature analogous to what we observed in Figure \ref{fig:varying_St}. This indicates that history effects contribute more significantly to the movement of the particle as its density diverges from neutrally buoyant. This effect can also be found in \citet{Prasath2019} Figure 5$(a)$, and may be especially applicable to cases where the density of the particle increases over time, such as with the effects of algal biofouling \citep{Kreczak2021}.

Additionally, we observe that simulations with history effects are consistently closer to the analytical solution for neutrally buoyant particles throughout the water column than simulations without history effects, similar to the observations made of Figure \ref{fig:varying_St}, where the drift velocities of particles simulated with history effects are closer to those of Stokes drag dominant or tracer-like particles. From our results thus far, we deduce that the history force influences microplastic particle transport more significantly as the size of the debris grows and as the density becomes very heavy or very light relative to the water, and that history effects slow particle relaxation.

\begin{figure}
    \centering
    \includegraphics{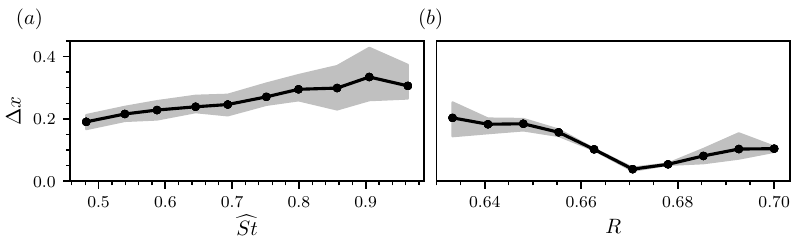}
    \caption{The difference in total horizontal displacement $\Delta x$ between a particle simulated with history effects and a particle simulated without history effects as a function of the Stokes number $(a)$ and the density ratio $(b)$. For all simulations in $(a)$, the density ratio is constant with $ R = 25 / 38 $, and similarly in $(b)$, the density-independent Stokes number is constant with $\widehat{St} = 0.34906$. Each point is an average value of $\Delta x$ taken over five simulation pairs, where a pair includes two simulations with identical conditions, differentiated by whether history effects are included. The shaded region indicates the maximum and minimum values computed from each set of five simulation pairs}
    \label{fig:displacement}
\end{figure}

Now, we examine the difference in the total horizontal displacement between a particle with history effects and a particle without history effects for various density ratios. To compute the difference in total horizontal displacement, we simulate the trajectory of a particle without history effects from an initial horizontal starting position $x_0$ until the particle either sinks to the seabed or rises to the water surface. Once the particle has reached the end of its trajectory, we record its final horizontal position $x_f$. Next, we run a second simulation for a particle under the same conditions, with the same initial horizontal position, and with history effects included. This provides $x_{f_h}$, the final horizontal position of the particle simulated with history effects. Then, the difference in total horizontal displacement $\Delta x$ may be computed as,
\begin{equation} \label{eq:displacement}
    \Delta x = \frac{\abs{x_{f_h} - x_f}}{\abs{x_f - x_0}}.
\end{equation}
To avoid the difference in horizontal displacement being influenced by the horizontal oscillation of the particle, for each variation in density, we choose five initial horizontal positions evenly spaced along one wavelength. From each of these initial horizontal positions $x_0$, we run a simulation with history effects and a simulation without history effects, and compute the difference in total horizontal displacement. The maximum displacement, minimum displacement, and mean displacement are then calculated from the solutions, as exhibited in Figure \ref{fig:displacement}.
We observe that particles close to neutrally buoyant have a low difference in horizontal displacement, while heavy or light particles exhibit a much larger mean difference. This is especially evident for negatively buoyant particles, indicating that the history force influences the transport of very heavy or very light particles far more strongly than those near neutral buoyancy. In general, the mean of $\Delta x$ is positive, meaning that history effects systematically enhance horizontal transport, and this enhancement is most pronounced at the lower end of $R$ considered here. As expected, $\Delta x$ generally increases with $St$ and fixed $R$, where mean transport increases by 20-30\% compared with simulations that do not include history in Figure~\ref{fig:displacement}a.  The spread around the mean is also significant. This variability is not numerical artefact, but a genuine dynamical effect and is important when comparing with experiments. In practice, experimental results could show large apparent scatter in horizontal transport that cannot be explained without incorporating the history force, a feature that earlier models were unable to capture. Accounting for this effect may therefore be important when interpreting observational or experimental data.
\subsection{Critical damping of the particle trajectory}
\begin{figure}
    \centering
    \includegraphics{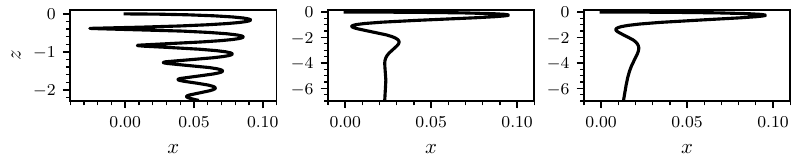}
    \caption{Negatively buoyant particle trajectories with $ R = 0.6 $, $ \epsilon = 2\upi / 75 $ exhibiting pre-critical (left, $ St = 0.5 $), critical (center, $ St = 4.0944 $), and post-critical (right, $ St = 10 $) oscillatory damping. All trajectories exclude history effects}
    \label{fig:critical_trajectories}
\end{figure}

As particle inertia increases (through larger size or greater density) the oscillations in their trajectories become progressively damped, as shown in Figure \ref{fig:critical_trajectories}. We previously observed how the history force slows particle relaxation across a range of sizes and densities (Figures \ref{fig:horizontal_forces} and \ref{fig:varying_St}-\ref{fig:varying_R}). Now, we examine whether the history force also influences the transition between oscillatory and overdamped motion. 

For our examination, we define a critically damped particle trajectory as a trajectory that has two or fewer period endpoints, where a period endpoint is determined according to the definition in \textsection \ref{subsec:param_variation}. Then, the critical Stokes number $ St_c $ corresponds to the maximum size of the particle before its trajectory is critically damped, for a fixed density. Similarly, the critical density ratio $ R_c $ is the minimum (in the case of negative buoyancy) or maximum (in the case of positive buoyancy) density ratio before the particle trajectory is critically damped, for a fixed size.
\begin{figure}
    \centering
    \includegraphics{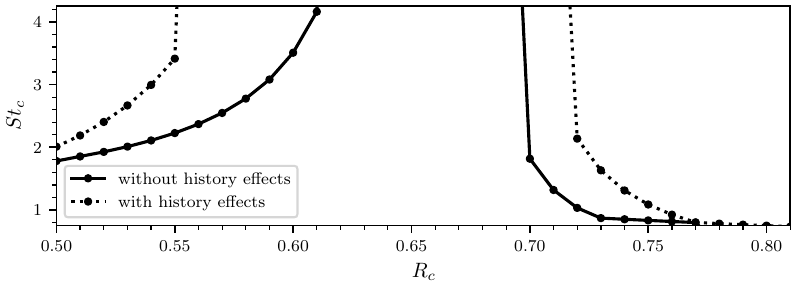}
    \caption{The critical Stokes number as a function of the critical density ratio for solutions with (dotted) and without (solid) history effects, where $ \epsilon = 2\upi / 75 $ for all simulations}
    \label{fig:critical_nums}
\end{figure}
In Figure \ref{fig:critical_nums}, we observe that when history effects are included, the Stokes number can be larger before its trajectory reaches critical damping, in comparison to a particle of the same density without history effects, and vice versa. This observation is consistent with several of our previous results: we consider Figures \ref{fig:horizontal_forces} and \ref{fig:varying_St}, which show that the influence of the history force on the horizontal drift of the particle increases as the Stokes number increases for a constant density ratio, along with Figure \ref{fig:varying_R}, which demonstrates that the influence of the history force increases as particle densities deviate from neutral buoyancy for a constant Stokes number. Further, taking into account the evidence that history effects slow particle settling shown in Figure \ref{fig:velocity_decay} as well as \citet{Prasath2019} Figure 5, we remark that large, heavy, or light particles simulated with history effects will settle much more slowly compared to identical particles simulated without history effects due to the greater influence of the history force for large Stokes numbers and non-neutral density ratios. Since particle trajectories exhibit increased oscillatory damping as they approach their terminal velocity, slow settling corresponds to a more oscillatory trajectory, and thus the inclusion of history effects decreases damping of the particle trajectory. For example, if a critically damped particle trajectory was simulated without history effects, an identical particle simulated with history effects will not exhibit a critically damped trajectory. However, it is important to recall that the Maxey-Riley equation necessitates $ \widehat{St} \ll 2/9 $, and thus the decreased damping effect of the history force is reduced within this range.
\subsection{Orbit shearing}
\begin{figure}
    \centering
    \includegraphics{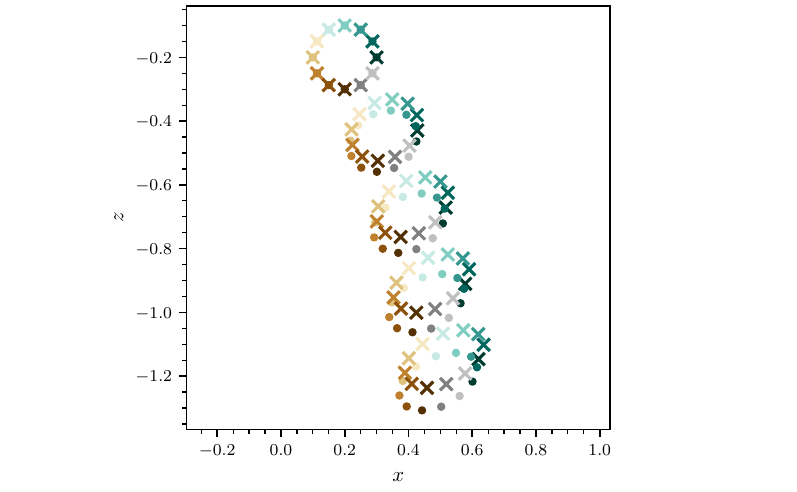}
    \caption{Snapshots of the trajectories of twelve negatively buoyant particles with history effects ($\times$) and without history effects ($\bullet$). For all particles, $ St = 0.127 $, $ R = 0.603 $, and $ \epsilon = \upi / 25 $. The position of each particle is shown every four wave periods}
    \label{fig:orbit_shearing}
\end{figure}

To further explore how the history force slows particle relaxation and settling, we investigate its influence on the dispersion of multiple particles. Here, we simulate the transport of a group of negatively buoyant particles released in a circular orientation near the top of the water column. As the particles oscillate and sink towards the seabed, the circular formation is distorted and becomes increasingly elliptical, illustrated in Figure \ref{fig:orbit_shearing}.
To quantify the shear $M$ of a particle within the circular formation, we take the $L^2$ norm of the composite matrix of horizontal and vertical shear,
\begin{equation} \label{eq:shear_matrix}
    M = \norm{\mqty{1 + m_x m_z & m_x \\ m_z & 1}}
\end{equation}
where $m_x$ and $m_z$ are the horizontal and vertical shear factors, respectively, computed as,
\begin{equation} \label{eq:shear_factor}
\begin{split}
    m_x &= \frac{x_f - x_0}{z_f}, \\ m_z &= \frac{z_f - z_0}{x_f}.
\end{split}
\end{equation}
with the initial position of the particle denoted $(x_0, z_0)$ and the final position of the particle denoted $(x_f, z_f)$. Then, to quantify the shearing of the circular formation itself, we compute the mean shear $\overline{M}$ of the particles constituting the shape. This provides a single solution for simulations with history effects, and one for simulations without. With these quantities, we can observe the significance of history effects on the shearing of the circular formation by taking the difference in mean shear $\Delta\overline{M}$ between simulations with and without history.

\begin{figure}
    \centering
    \includegraphics{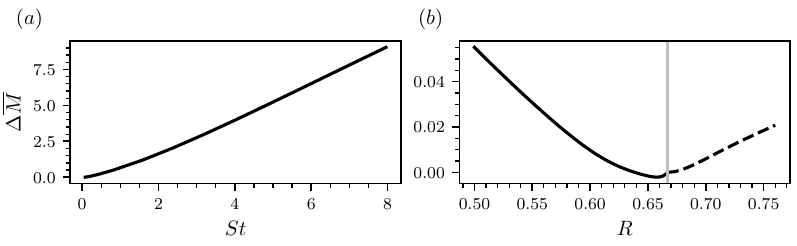}
    \caption{The difference in shear between simulations run with and without history effects for varying Stokes numbers $(a)$ and particle densities $(b)$. In $(a)$, the density ratio is constant, with $ R = 0.60\bar{3} $, and in $(b)$, the density-independent Stokes number is constant, with $\widehat{St} = 0.12$. The vertical line in $(b)$ separates negatively buoyant (solid) from positively buoyant (dashed) particles, and denotes neutral buoyancy. For all simulations, $ \epsilon = \upi / 25 $}
    \label{fig:shear_difference}
\end{figure}

Figure \ref{fig:shear_difference} exhibits the mean shear difference for various particle sizes and densities, achieved through variation in the Stokes number $(a)$ and density ratio $(b)$. The mean shear difference in $(a)$ is entirely positive and increases steadily as the Stokes number grows and the density ratio remains constant ($R = 0.60\bar{3}$), indicating that the circular formation comprising particles simulated with history effects undergoes more significant shearing than its counterpart omitting history effects. This effect is amplified for larger particles, a characteristic analogous to our observations of the increased influence of the history force for particles with larger Stokes numbers in Figures \ref{fig:horizontal_forces} and \ref{fig:varying_St}.

Similarly, with a constant density-independent Stokes number ($\widehat{St} = 0.12$), the mean shear difference in $(b)$ decreases for negatively buoyant particles as their density approaches that of the fluid, reaching zero at neutral buoyancy. Then, as buoyancy increases, the mean shear difference steadily grows. While the difference in mean shear is entirely positive for positively buoyant particles, there is a region at approximately $ 0.63 < R < 2/3 $ where the mean shear difference is negative, meaning that for slightly negatively buoyant particles, the circular formation composed of particles simulated without history effects experienced more shearing than the formation with history effects. The physical justification of this feature is not clear, and has potential for future investigation. The regions where the mean shear difference is largest, on the other hand, suggest that the inclusion of the history force contributes significantly to increased particle dispersion. These regions are composed of very heavy and very light particles relative to the fluid, and the observation that the influence of the history force is stronger for these particles is substantiated by our previous findings in Figures \ref{fig:varying_R} and \ref{fig:displacement}.
\section{Discussion \& concluding remarks} \label{sec:conclusion}
In this paper, we have provided a detailed examination of how memory effects affect the motion of small inertial particles in unsteady surface wave flows. While prior studies have examined the role of the history force in simplified sedimentation or oscillatory settings, we have analyzed this force within a physical context representative of microplastic transport. We provide a framework for interpreting particle-wave interactions by analyzing the force balance throughout the particle's trajectory and quantifying the role of history effects across a wide parametric range. We clarify how non-local hydrodynamic effects, which are often neglected in wave-particle modeling, introduce significant phase shifts, alter relaxation dynamics, influence residence time in the wave-affected region, and significantly enhance horizontal transport. 

A key outcome of our analysis is the identification of a physically interpretable parameter, $S=St/\gamma^2$, which approximately collapses the behavior across particle densities and provides a sensible threshold for when history effects become dynamically important. We find that memory effects overtake Stokes drag once $S$ exceeds a critical value $S\approx 0.25$: this corresponds to $\widehat{St}^*\approx0.25$, which is almost a factor of three smaller than one would expect from a standard inertial approximation. This allowed us to construct the first regime map (Figure \ref{fig:parameter_map}) that delineates three distinct transport regimes: a Stokes-drag regime ($S<0.0025$), a non-negligible history regime ($0.0025\leq S\leq 0.25$), and a history-dominant regime ($0.25< S$). For representative sea states, we found that even moderately sized polystyrene microplastics fall in the parameter range where history effects are non-negligible ($a'\geq 0.14\,$mm) or dominant ($a'>1.4\,$mm). 

Overall, our results indicated that the inclusion of history effects is necessary for models attempting to simulate the transport of particles within the full range of microplastic particle sizes. We further demonstrated that history effects substantially increase horizontal transport distances compared with models that neglect them, and that they enhance orbit shearing within particle ensembles. Indeed, the effects of the history force become even more important for the movement of particles that are very large, very heavy, or very light with respect to the water density. History effects also generate a pronounced spread in transport distances that is absent from models using instantaneous drag alone, especially for negatively buoyant particles. This variability is an intrinsic dynamical feature and is therefore crucial to consider when interpreting observational or experimental measurements of microplastic transport.

Like modifications to the Stokes drag term \citep{Dibenedetto2022} or the inclusion of the history force for settling or oscillatory flows \citep{Prasath2019}, our implementation of the history force for particle transport in linear waves improves the accuracy of theoretical models for microplastic particle transport in ocean waves. A potential extension to this application would be to include modifications to the Stokes drag term, as in \citet{Dibenedetto2022}, along with our implementation of the history force, to allow for both larger particle sizes and particle Reynolds numbers in the transitional range. Additionally, improvements could be made to the computational efficiency of the implementation, providing faster execution of simulations with wavelengths in the full range described in Table \ref{tab:dimensional_parameters}, as simulations with long wavelengths are among the most restrictive for our current implementation. While the scope of this paper is focused on linear waves, this implementation could be adapted to other, more complicated flows as well. Expanding the application of the model to non-linear wavy flows would allow for steeper waves, thus increasing the applicable parametric ranges of the model. An investigation of non-linear flows should include a determination of the boundary between the Stokes drag-dominant, non-negligible history, and history dominant regimes. To further increase its robustness, the model could be modified to be applicable for flows composed of discrete experimental data. This would allow for applications to flow fields generated from computational fluid dynamics software, such as OpenFOAM or SWASH, rather than being restricted to flows with exact mathematical expressions.

%\backsection[Supplementary data]{\label{SupMat}Supplementary material and movies are available at \\https://doi.org/10.1017/jfm.2019...}

%\backsection[Acknowledgments]{We thank Zhiling Liao, Qingping Zou and Jacques Vanneste for useful discussions.}

%\backsection[Funding]{This research was supported by the UK Engineering and Physical Sciences Research Council (EPSRC) through the MAC-MIGS Centre for Doctoral Training (EP/S023291/1). }

\backsection[Declaration of interests]{The authors report no conflict of interest.}

%\backsection[Data availability statement]{The data that support the findings of this study are openly available in [repository name] at http://doi.org/[doi], reference number [reference number]. See JFM's \href{https://www.cambridge.org/core/journals/journal-of-fluid-mechanics/information/journal-policies/research-transparency}{research transparency policy} for more information}

%\backsection[Author ORCIDs]{ C. Cummins, https://orcid.org/0000-0001-7091-37; B. Jones, https://orcid.org/0000-0009-8765-4321}

%\backsection[Author contributions]{Authors may include details of the contributions made by each author to the manuscript'}
%\appendix
%\section{}\label{appA}
\newpage
\renewcommand{\thefigure}{A\arabic{figure}}
\bibliographystyle{jfm}
\bibliography{references}
\end{document}